**Bosonic phases across the superconductor-insulator transition in infinite-layer samarium nickelate**

Menghan Liao,[1,#,*] Heng Wang,[2,3,#] Mingwei Yang,[4,5,#] Chuanwu Cao,[1,#] Jiayin Tang,[4] Wenjing Xu,[4] Xianfeng Wu,[3] Guangdi Zhou,[2,3] Haoliang Huang,[2,3] Kaiwei Chen,[1,6,7] Yuying Zhu,[1,8] Peng Deng,[1] Jianhao Chen,[1,8,9,10] Zhuoyu Chen,[2,3,†] Danfeng Li,[4,5,‡] Kai Chang,[1,§] Qi-Kun Xue[1,2,3,11]

[1] *Beijing Key Laboratory of Fault-Tolerant Quantum Computing, Beijing Academy of Quantum Information Sciences, Beijing 100193, China*

[2] *Quantum Science Center of Guangdong-Hong Kong-Macao Greater Bay Area, Shenzhen 518045, China.*

[3] *State Key Laboratory of Quantum Functional Materials, Department of Physics, and Guangdong Basic Research Center of Excellence for Quantum Science, Southern University of Science and Technology, Shenzhen 518055, China*

[4] *Department of Physics, City University of Hong Kong, Kowloon, Hong Kong SAR, China.*

[5] *Shenzhen Research Institute of City University of Hong Kong, Shenzhen 518057, China.*

[6] *Beijing National Laboratory for Condensed Matter Physics, Institute of Physics, Chinese Academy of Sciences; Beijing 100190, China.*

[7] *University of Chinese Academy of Sciences, Beijing 100049, China.*

[8] *Hefei National Laboratory, Hefei 230088, China*

[9] *International Center for Quantum Materials, School of Physics, Peking University, Beijing, 100871 China.*

[10] *Key Laboratory for the Physics and Chemistry of Nanodevices, Peking University, Beijing, 100871 China.*

[11] *State Key Laboratory of Low Dimensional Quantum Physics and Department of Physics, Tsinghua University, Beijing, 100084, China*

[#]these authors contribute equally

**ABSTRACT.** Superconductivity arises from the global phase coherence of Cooper pairs. Modulation of phase coherence leads to quantum phase transitions, serving as an important tool for studying unconventional superconductivity. Here, we demonstrate bosonic phases across the superconductor-insulator transition in infinite-layer nickelate superconducting films by the control of spatially periodic network patterns. Magnetoresistance oscillations with a periodicity of $h/2e$ provide direct evidence of $2e$ Cooper pairing in nickelates. The phase transition is predominantly driven by enhanced superconducting fluctuations, and Cooper pairs are involved in charge transport across the transition. Notably, we observe two types of anomalous metallic phases, emerging respectively at finite magnetic fields and down to zero magnetic field. They can be characterized by bosonic excitations, suggesting the dynamic roles of vortices in the ground states. Our work establishes nickelates as a key platform for investigating the rich landscape of bosonic phases controlled via the phase coherence of Cooper pairs.

Correspondence:
*liaomh@baqis.ac.cn
†chenzhuoyu@sustech.edu.cn
‡danfeng.li@cityu.edu.hk
§changkai@baqis.ac.cn

## I. INTRODUCTION

In Bardeen–Cooper–Schrieffer (BCS) superconductors, electrons near the Fermi surface form Cooper pairs and condense into a phase-coherent ground state [1]. Breaking the Cooper pairs or the phase coherence, which occur simultaneously in conventional superconductors, suppresses superconductivity [2]. Cooper pairing without establishing global phase coherence gives rise to exotic physical phenomena [3-7]. For instance, in two-dimensional (2D) superconductors or Josephson junction arrays (JJA), the dissipationless state can emerge via a Berezinskii-Kosterlitz-Thouless (BKT) transition at a temperature significantly lower than the Cooper pairing temperature [3-5]. This is because the unbinding of vortex-antivortex pairs below the pairing temperature disrupts the phase coherence. In cuprate high-temperature superconductors, it is believed that the superconducting critical temperature is determined by the onset of phase coherence, while Cooper pairs may form at much higher temperatures [6, 7]. This is due to the fact that the superfluid density is relatively low, and phase fluctuations are important. The separation between electron pairing and phase coherence is a possible explanation for enduring mysteries in high-temperature superconductivity, such as the origin of the pseudogap [8].

Nickelate superconductors are a new class of high-temperature superconductors. Superconductivity in nickelates was first observed in $Nd_{0.8}Sr_{0.2}NiO_2/SrTiO_3$ with critical temperatures ($T_c$) ranging from 9 to 15 K [9]. The $T_c$ of Ruddlesden-Popper nickelates has recently surpassed the boiling point of liquid nitrogen under high pressure [10] and the McMillan limit under ambient pressure [11, 12]. However, the superconducting transition in most nickelates remains notably broad, with zero resistance achieved significantly below the superconducting onset temperature ($T_{c,\ onset}$). Among the nickelate superconductors, the infinite-layer nickelates ($RNiO_2$, where R represents rare-earth elements) are of particular interest due to their structural similarity to infinite-layer cuprates [13]. They share a $3d^9$ electron configuration, and the electronic structures near the Fermi level are analogous in some respects. Their resemblance has led to the proposal of possible similar superconducting properties. As such, $RNiO_2$ offers a valuable platform to test theories of high-temperature superconductivity. For instance, both $RNiO_2$ and cuprates seem to display a significantly lower superfluid density compared to conventional superconductors [6,7,14-16], suggesting that they are vulnerable to phase fluctuations of the Cooper pairs. Nevertheless, the role of superconducting fluctuations in nickelate superconductors has been little investigated experimentally and remains an open question.

In this work, we manipulate the behavior of Cooper pairs in $Sm_{0.95-x}Eu_xCa_{0.05}NiO_2$ films by controlling the phase coherence via patterning the films into spatially periodic networks. We realize a superconductor-insulator transition (SIT). Magnetoresistance (MR) oscillations with a period of $h/2e$ reveal that Cooper pairs participate in the electrical transport at low temperatures even when dissipationless transport is suppressed. Enhanced superconducting fluctuations modulate the transport properties of the Cooper pairs significantly and lead to a series of bosonic phases. We observe two types of anomalous metallic (AM) states, characterized by an unusual saturation of resistance to finite values as temperature approaches zero. The first type is observed when the applied magnetic field disrupts the phase coherence transport of the Cooper pairs. With increased superconducting fluctuations, we observed a second type of AM state in the absence of an external magnetic field in the intermediate state of the SIT. It arises upon cooling from a bosonic strange metal state, characterized by linearly temperature-dependent resistance ($R_s(T)$) in a wide temperature range. When superconducting fluctuations are strong enough, the system finally becomes a Cooper pair insulator. Our work highlights how superconducting fluctuations control the bosonic transport and give rise to intriguing physical phenomena in nickelate superconductors.

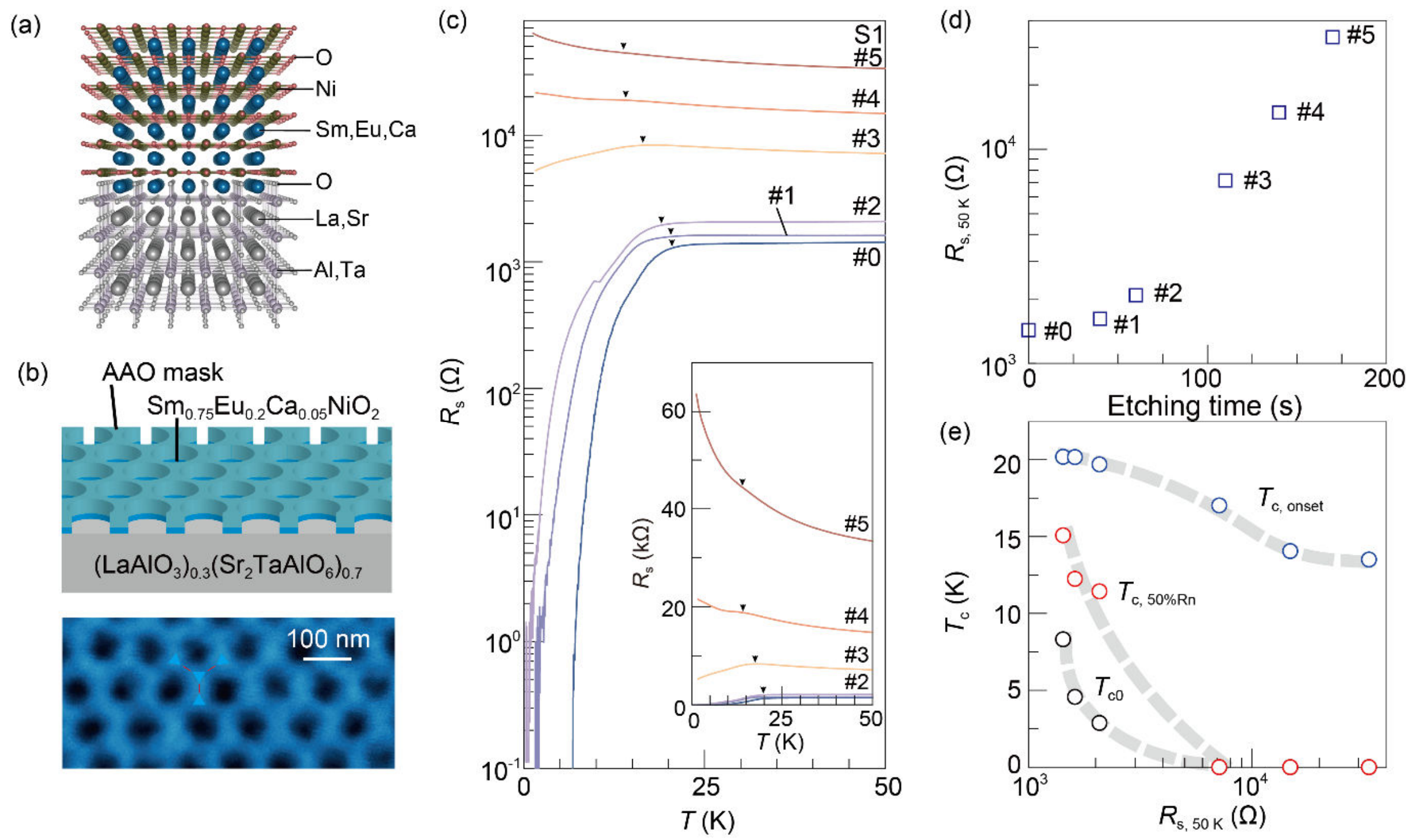

FIG. 1. Temperature dependent resistance ($R_s(T)$) measurements of $Sm_{0.95-x}Eu_xCa_{0.05}NiO_2$ at different etching steps (sample S1). (a) Crystal structure of pristine $Sm_{0.95-x}Eu_xCa_{0.05}NiO_2$ on $(LaAlO_3)_{0.3}(Sr_2TaAlO_6)_{0.7}$ substrates. (b) Top panel: Schematic illustration of the $Sm_{0.95-x}Eu_xCa_{0.05}NiO_2$ network fabricated by RIE through AAO masks. Bottom panel: SEM image of the networks. Blue triangles indicate superconducting islands, and red lines denote weak links between them. (c) $R_s(T)$ at each etching step. The numbers of the steps are marked next to the corresponding curves. Black triangles mark the kinks in the curves ($T_{c,\ onset}$). The inset shows the same data on a linear scale. (d) Sheet resistance at 50 K ($R_{s,\ 50\ K}$) as a function of total etching time, with step numbers labeled next to the blue rectangles. (e) Evolution of superconducting transition temperatures with $R_{s,\ 50\ K}$. The blue, red, and black dots represent $T_{c,\ onset}$, $T_{c,\ 50\%Rn}$ and $T_{c0}$ as a function of $R_{s,\ 50\ K}$, respectively. Gray dashed curves are guides to the eye.

## II. RESULTS

### A. SIT in samarium nickelate

To fabricate nanopatterned nickelate networks, we prepare ~9 nm-thick $Sm_{0.95-x}Eu_xCa_{0.05}NiO_2$ films (with $x = 0.24$, 0.2 for sample S1 and S2, respectively, and $x = 0.22$ for sample S3 and S4) on $(LaAlO_3)_{0.3}(Sr_2TaAlO_6)_{0.7}$ substrates using pulsed laser deposition (PLD) and topotactic reduction. The lattice structure of the as-grown film is shown in Fig. 1(a). The superconducting properties of the pristine films have been previously reported [17, 18]. Anodized aluminum oxide (AAO) masks with a triangular array of holes (50 nm hole diameter, 100 nm center-to-center distance) are transferred onto the films using a wet transfer process. In Fig. S7 in the Supplemental Material [19], we show that the films covered with AAO masks exhibit little change in the superconducting transition temperature compared with bare samples. Reactive ion etching (RIE) is then performed to pattern the nickelate films. The AAO mask remains on the film during subsequent electrical transport measurements. Fig. 1(b) shows a schematic illustration and a scanning electron microscopy (SEM) image of the nanopatterned films. More details on the film growth, topotactic reduction, and device fabrication are provided in the Methods section.

Electrical transport measurements are carried out as follows: we etch the films for a certain period of time, then load them into the cryostat for measurements. Afterward, the sample is unloaded, re-etched, and measured again. By repeating this process multiple times, we obtain a series of transport data corresponding to different etching steps of sample S1 (labeled S1#0 to S1#5, with S1#0 referring to the pristine film). Fig. 1(c) displays $R_s(T)$ at each etching step. S1#0 undergoes a superconducting transition with $T_{c,\ onset}$ around 20 K, and reaches zero resistance at $T_{c0} \approx 7.5$ K. As the total etching time increases, the resistance rises (Fig. 1(d)), while the superconductivity is suppressed (Fig. 1(c)). S1#3 exhibits a metallic behavior below $T_{c,\ onset}$. Starting from S1#4, the film enters the insulating state. The SIT is also reproduced in sample S3 with a different doping level (Fig. S7-S12 in the Supplemental Material). The SIT can be attributed to the increasing disorder level introduced by RIE along the hole sidewalls [20]. As etching proceeds, the disordered regions become more pronounced, and the continuous superconducting film breaks into superconducting islands (the wider regions of the film, marked by the blue triangles in Fig. 1(b)) connected by weak links (the narrower regions of the film, shown by the red lines in Fig. 1(b)). X-ray diffraction (XRD) measurements performed after etching indicate that the crystal structure of the superconducting islands is little affected (See Fig. S13 in the Supplemental Material). In S1#3-#5 where global phase coherence is lost and zero resistance is not observed, the kinks (marked by the arrows in Fig. 1(c)) around 15 K~20 K in each $R_s(T)$ curve indicate the presence of Cooper pairing within the superconducting islands.

As total etching time increases, the superconducting transition becomes broader. This is more clearly illustrated in Fig. 1(e), where we plot $T_{c,\ onset}$, $T_{c,\ 50\%Rn}$, and $T_{c0}$, corresponding to the temperatures at which kinks appear in $R_s(T)$, and where $R_s$ drops to 50% and 1% of the normal-state resistance at 25 K ($R_n$), respectively. As the sheet resistance at 50 K ($R_{s,\ 50\ K}$) increases, both $T_{c0}$ and $T_{c,\ 50\%Rn}$ rapidly drop to zero. In contrast, $T_{c,\ onset}$ experiences a mild decrease from 20 K to 15 K, although $R_{s,\ 50\ K}$ increases by a factor of 20. These observations further confirm that the etching primarily weakens the coupling between superconducting islands, while the pairing strength within the islands is less affected. The broadening of the superconducting transition signals the enhanced superconducting fluctuations at higher disorder levels [21].

### B. MR oscillations with a periodicity of *h*/*2e*

Having shown that Cooper pairing persists across the SIT, we next investigate the magnetic field dependent resistance at low temperatures. As shown in Fig. 2(a), the pristine sample (S1#0) maintains zero resistance below 0.8 T (resistance measurements up to 9 T of S1#0 and S3#0 are plotted in Fig. S1 and Fig. S9 in the Supplemental Material, respectively). In S1#2, a small magnetic field suppresses the dissipationless transport.

Remarkably, resistance oscillations (marked by triangles in Fig. 2(a)) are observed in S1#3, where finite resistance is observed even in the absence of a magnetic field. The oscillations have a period of approximately 0.23 T, remaining visible in the insulating regime (S1#4). In sample S2, which is similarly patterned by etching through an AAO mask (S2#6), up to three oscillation periods are observed (Fig. 2(b), $R_s(T)$ of sample S2 at different etching steps is shown in Fig. S5 in the Supplemental Material). To better visualize the oscillations, we subtract a background from the MR (MR $= R(\mu_0 H)/R(0) - 1$) and plot the resulting δMR as a function of the magnetic field (bottom panel of Fig. 2(b); details of the background subtraction procedure are provided in Fig. S6 in the Supplemental Material). Fig. 2(c) plots MR and δMR of S1#3. As indicated by the gray dashed lines in Fig. 2(b) and (c), the oscillation period corresponds to one superconducting flux quantum ($\phi_0$ = $h/2e$) per unit cell of the nickelate network's honeycomb lattice (with an area of ~8660 nm²). Fig. S2 in the Supplemental Material shows that the oscillations are independent of the excitation current $I$ when $I \leq 200$ nA. The periodicity of $h/2e$ is unambiguous evidence that Cooper pairs are involved in the electrical transport, despite the absence of zero resistance [22-28].

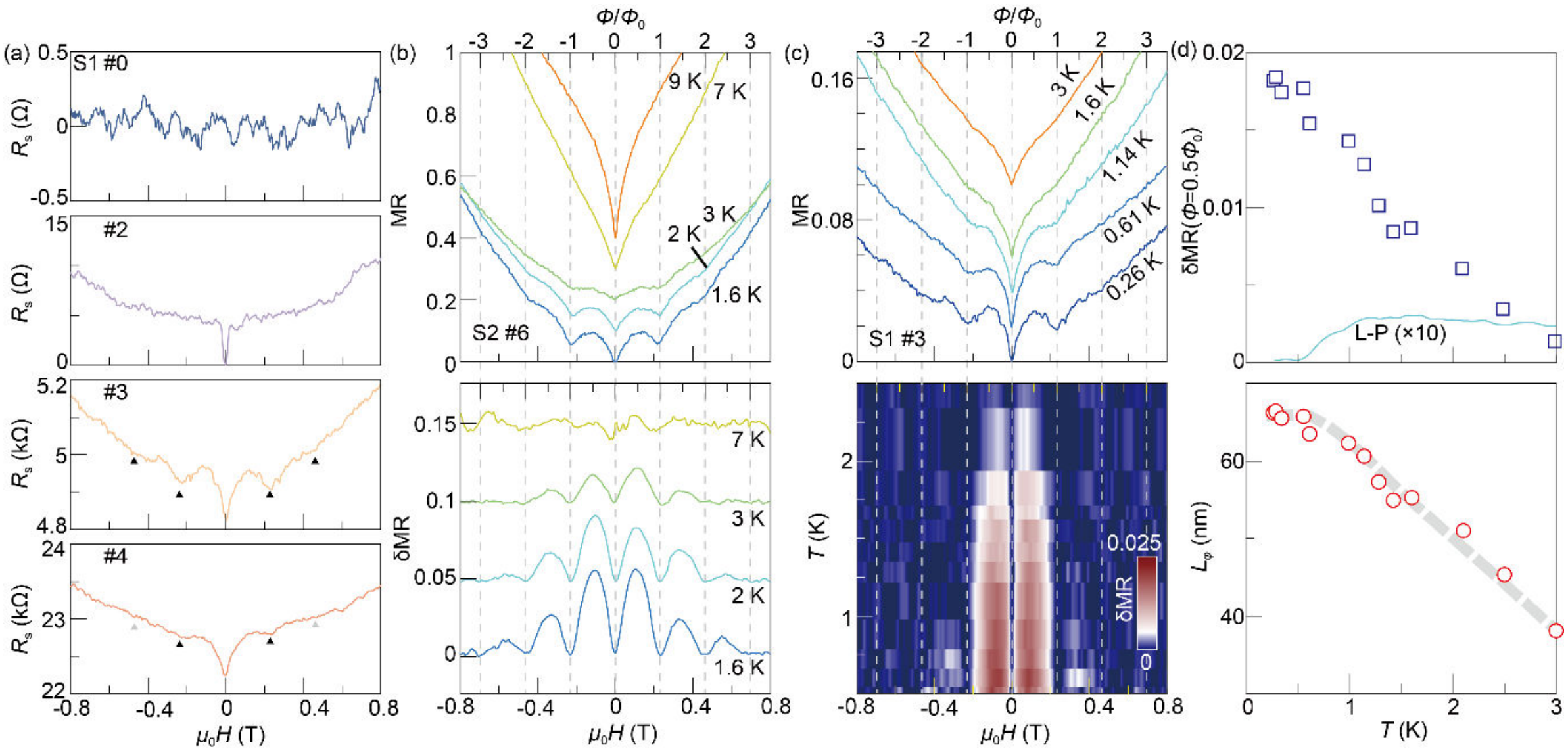

FIG. 2. Magnetoresistance (MR) oscillations with a period of $h/2e$ in the $Sm_{0.95-x}Eu_xCa_{0.05}NiO_2$ networks. (a) Sheet resistance as a function of applied magnetic field of sample S1 at different etching steps (labeled in each panel). Data of S1#0 were recorded at 1.6 K, and data of S1#2-#4 were taken at 0.26 K. Black and gray triangles mark the resistance dips associated with oscillations. (b) Top panel: MR of S2#6 at different temperatures. Bottom panel: Background-subtracted magnetoresistance (δMR) at different temperatures. The temperatures are labeled next to the curves. Gray dashed lines indicate magnetic field values corresponding to integer numbers of flux quanta ($\Phi = n\times\Phi_0$) through one unit cell of the superconducting network. (c) Temperature dependent MR (top panel) and δMR (bottom panel) of S1#3. (d) Top panel: Blue rectangles represent the oscillation amplitude at $\Phi=0.5\Phi_0$ of S1#3 at different temperatures extracted from the color plot in (c). The blue curve shows the oscillation amplitude as a function of temperature for the Little-Parks effect, scaled by a factor of 10 for comparison. Bottom panel: Red dots show the temperature dependent coherence length estimated from the oscillation amplitude in the top panel. The gray dashed curve is a guide to the eye.

To investigate the mechanism of the oscillations, we study the temperature dependence of the oscillation amplitude. The blue squares in Fig. 2(d) show the oscillation amplitude at $\phi = 0.5\phi_0$ as a function of temperature extracted from the color plot in Fig. 2(c). The amplitude increases when cooling down from 3 K and saturates below 0.5 K. From the oscillation amplitude, we estimate the phase coherence length $L_\phi$ of the charge carriers using the expression $\delta\mathrm{MR}(\phi = 0.5\phi_0) = R_s\frac{4e^2}{h}(\frac{L_\phi}{\pi r})^{1.5}\exp(-\frac{\pi r}{L_\phi})$ [24, 26, 29], where $r\sim50$ nm is half of the center-to-center distance between the neighboring holes. As plotted in the bottom panel of Fig. 2(d), the phase coherence length saturates at approximately 65 nm below 0.5 K. The oscillation amplitude in the insulating state (S1#4) at base temperature reduces to $\delta\mathrm{MR}(\phi = 0.5\phi_0) = 0.009$, corresponding to a coherence length of 40.3 nm. This is possibly because the stronger disorder level weakens the phase coherence transport of the charge carriers. A similar drop in phase coherence length as the etching time increases is observed in sample S3 (Fig. S10 in the Supplemental Material). In periodically patterned superconducting networks, resistance oscillations are typically attributed to two mechanisms: (1) the Little-Parks effect [30], (2) the formation of a JJA [31]. The oscillation amplitude associated with the Little-Parks effect is described by the expression $\delta\mathrm{MR}(\phi = 0.5\phi_0) = 0.14\frac{T_{c,\mathrm{onset}}}{R_S} \times \left(\frac{\xi_0}{r}\right)^2 \times \frac{dR_S}{dT}$, where $\xi_0 \sim 2$ nm is the superconducting coherence length [18], $dR_S/dT$ and $T_{\mathrm{c,\ onset}}$ are obtained from the $R_{\mathrm{s}}(T)$. The blue curve in the upper panel of Fig. 2(d) shows the predicted temperature dependent oscillation amplitude. Evidently, the observed oscillation amplitude is nearly two orders of magnitude larger and exhibits a significantly different temperature dependence, ruling out the Little-Parks mechanism. Instead, the oscillations here resemble those of JJA [31]. In the dissipative regime, the tunneling rate of Cooper pairs, which affects the resistance, is periodically modulated by the applied magnetic field, resulting in resistance oscillations [22, 24, 31].

### C. AM and bosonic strange metal states

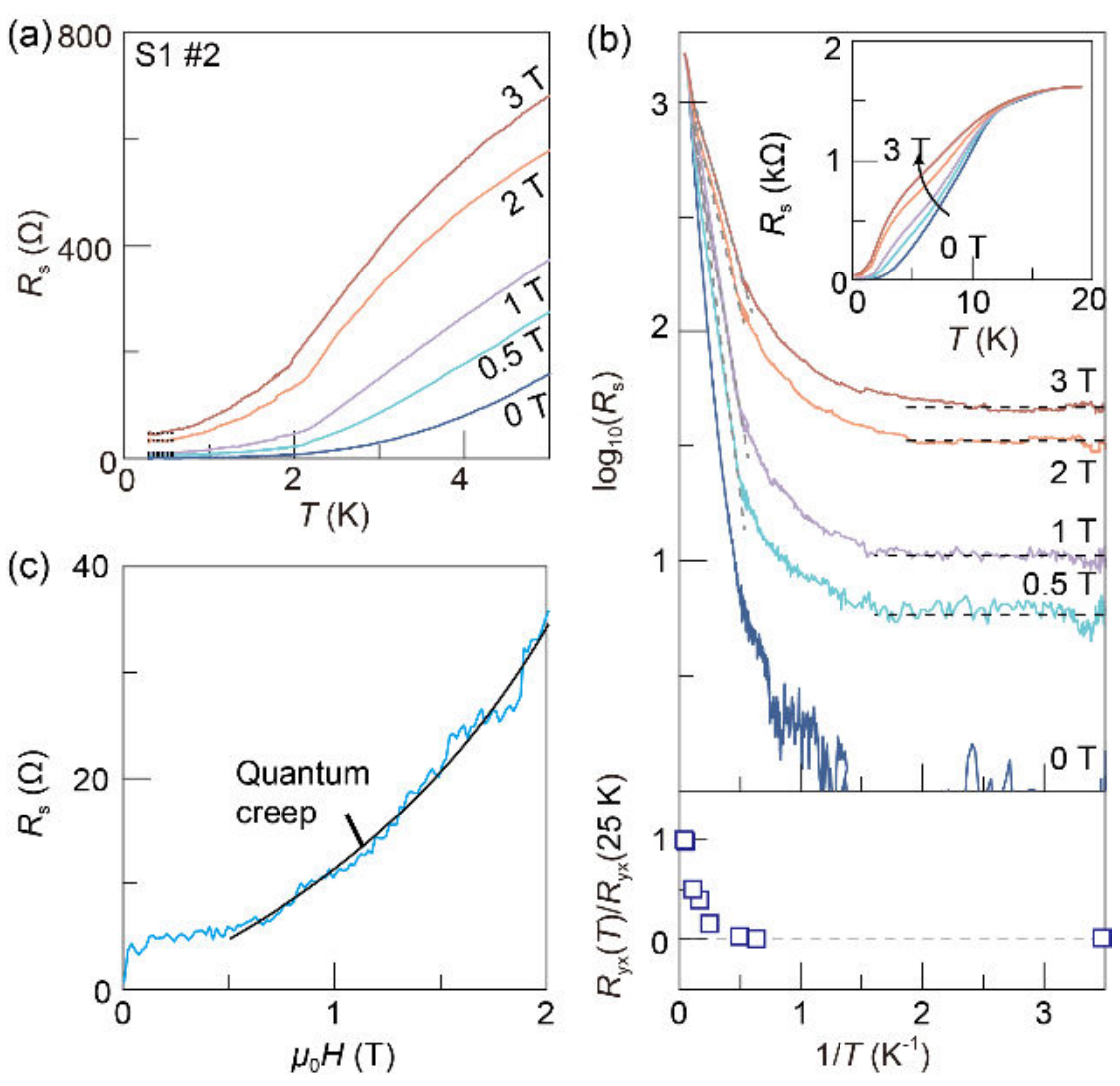

FIG. 3. The anomalous metallic (AM) state under finite magnetic fields in S1#2. Panel (a) and the inset of (b) show $R_{\mathrm{s}}(T)$ at different applied magnetic fields on a linear scale. The upper panel of (b) presents the same data in an Arrhenius plot. Magnetic field values are labeled next to each curve. Black dashed lines highlight the saturation of resistance to finite values at low temperatures, indicating the AM state. In the upper panel of (b), gray lines represent linear fits based on the thermally activated flux flow model. The lower panel of (b) shows the Hall coefficient normalized to its value above $T_{\mathrm{c,\ onset}}$ at 25 K. Data are obtained at ±8.5 T. (c) The blue curve shows the sheet resistance as a function of applied magnetic field at 0.26 K. The black curve is a fit using the quantum creep model in the magnetic field range of 0.5 T to 2 T.

After demonstrating that Cooper pairs contribute to transport across the SIT and that etching enhances superconducting fluctuations, we investigate the fluctuation-affected transport behavior of S1#2 and S1#3 by performing temperature-dependent resistance measurements under various magnetic fields. Fig. 3(a) and (b) show the data of S1#2. As indicated by the gray dashed lines in the upper panel of Fig. 3(b), the $R_{\mathrm{s}}(T)$ at high temperatures below $T_{\mathrm{c,\ onset}}$ follows a linear behavior in the Arrhenius plot. This suggests that the resistance is caused by the thermally assisted collective vortex creep [32], and the resistance can be described by $R_S \sim \exp\left[U(\mu_0 H)/T\right]$, where $U(\mu_0 H)$ is the magnetic field dependent activation energy [32] (the extracted $U(\mu_0 H)$ is plotted in Fig. S3 in the Supplemental Material). However, as the temperature approaches zero, the resistance deviates from the thermal activation behavior and saturates at a finite value. This is the hallmark of the AM state [33], observed in various 2D superconductors and violating the expectation of Anderson localization [34]. The mechanism of the resistance saturation—sometimes to a value significantly lower than predicted by the Drude model [35]—remains one of the long-standing mysteries in 2D superconductivity [33]. In Note 1 of the Supplemental Material, we show that the resistance saturation is independent of the excitation current and the use of RC filters, which demonstrates that the AM state is intrinsic. We also discuss that classical percolation fails to explain the resistance saturation. We observe suppression of the Hall signal at low temperatures (lower panel of Fig. 3(b)), a characteristic feature of the AM state [24, 33, 36]. One of the possible origins of this AM state is the quantum creep of vortices generated by an external magnetic field [37, 38]. In this model, the quantum motion of vortices leads to fluctuations in the superconductivity down to zero temperature, giving rise to a finite resistance [37, 38]. Indeed, the magnetic field dependent resistance across a broad field range where the AM state is observed can be fitted by the quantum creep model [38]: $R_s \sim \frac{h}{4e^2}\frac{\kappa}{1-\kappa}, \kappa \sim \exp\left[C\left(\frac{H-H_0}{H}\right)\right]$ (Fig. 3(c)), where C and $H_0$ are the fitting parameters, suggesting that magnetic field induced quantum creep of vortices is one of the possible mechanisms for the AM state observed in S1#2.

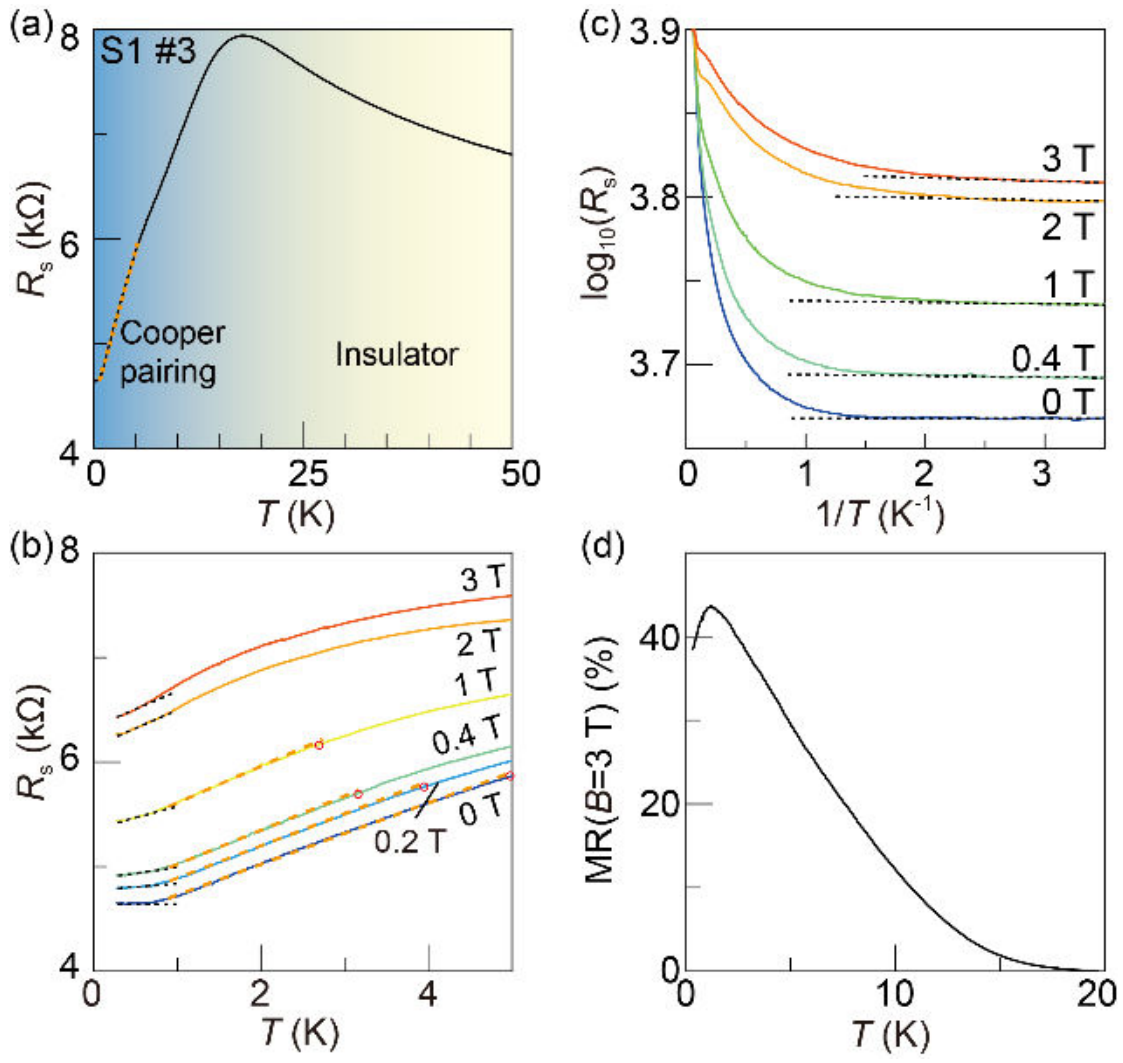

FIG. 4. The AM state and bosonic strange metal state in S1#3. (a) $R_s(T)$ at zero magnetic field. The yellow and blue regions show the insulating behavior above $T_{c\ onset}$ and the metallic behavior due to Cooper pairing below $T_{c\ onset}$, respectively. (b), (c) $R_s(T)$ measured under different magnetic fields, shown on a linear scale (b) and in the Arrhenius plot (c). The orange dashed lines in (a) and (b) highlight the regime of the nearly linear $R_s(T)$. The red circles in (b) mark the onset temperatures of the strange metal state at different magnetic fields, which are defined as the $R_s(T)$ deviating by 0.3% from the linear fits (orange dashed lines). The black dashed lines in (b) and (c) indicate the resistance saturation under zero magnetic field and the suppression of the resistance saturation at finite magnetic fields. (d) Magnetoresistance at 3 T as a function of temperature.

As the disorder strength is further increased, the $R_s(T)$ of S1#3 above $T_{c,\ onset}$ exhibits insulating behavior, highlighted by the yellow region in Fig. 4(a). Below $T_{c,\ onset}$, Cooper pairs begin to form, and the resistance drops but does not reach zero. Between 5 K and 1 K —a temperature range spanning a factor of five— the resistance exhibits a nearly linear temperature dependence at zero applied magnetic field (marked by the orange dashed lines in Fig. 4(a) and (b)). The linear dependence is observable below 1 T, although the onset temperature of the linear regime (shown by red circles in Fig. 4(b)) decreases at higher fields. The linear $R_s(T)$ over a wide temperature range and persisting down to very low temperatures, referred to as a strange metal, has been observed in multiple strongly correlated systems [25, 26, 39-42]. This phenomenon defies the predictions of conventional Fermi liquid theory and remains one of the very important unresolved questions in condensed matter physics. Recently, the strange metallicity has been observed in bosonic metallic states such as superconducting networks patterned from cuprate and Fe-based high-temperature superconducting films [25, 26], while it is absent in nanopatterned $LaAlO_3/KTaO_3$ interface superconductors [27]. Key characteristics include linear $R_s(T)$ below $T_{c,\ onset}$ and resistance oscillations with a period of $h/2e$ [25, 26]. Our observation of bosonic strange metal states in nanopatterned nickelate superconducting films suggests that it is likely a universal phenomenon of nanopatterned high-temperature superconductors.

Notably, S1#3 transitions into an AM state in the absence of the magnetic field, indicated by a resistance saturation below 0.7 K (the bottom curves in Fig. 4(b) and (c)). Interestingly, applying a magnetic field suppresses the resistance saturation (marked by the black dashed lines in Fig. 4(b) and (c)), suggesting weakening of the AM state. This could be attributed to enhanced thermal excitation of vortex motion as the vortex density induced by the external magnetic field increases [ 43 ]. Importantly, the disappearance of resistance saturation under an applied magnetic field rules out extrinsic factors — such as Joule heating— as its origin. If extrinsic mechanisms were responsible, the saturation would persist regardless of the magnetic field. This observation further supports the intrinsic nature of the AM state. The AM state persisting down to zero magnetic field exhibits giant positive MR [24, 33]. Fig. 4(d) shows the temperature dependent MR at 3 T of S1#3, which exhibits a sharp increase below 20 K and reaches 40 % at low temperatures. Fig. S12 in the Supplemental Material shows that the positive MR in S3#2 can reach a giant value of 780% [24]. In S1#3, the AM state is observed without an applied magnetic field, implying a mechanism distinct from the quantum vortex motion driven by external fields in S1#2. The transition from bosonic strange metal at high temperatures to the AM state when the temperature approaches zero in a 2D superconducting system has been attributed to Ohmic dissipation of phase fluctuations, caused by coupling between bosonic and fermionic modes [26]. A linear $R_s(T)$ is expected when the resistance of the weak links $R_L$ approaches $R_c = \frac{h}{4e^2}$ ($R_L$ is close to $R_n$, assuming the sample in the normal state can be modeled as an infinitely large resistor network with honeycomb geometry of nodes and the resistance between the adjacent nodes is $R_L$ [44]). Indeed, the $R_n$ of S1#3 is close to $R_c$, suggesting that the AM and bosonic strange metal state in S1#3 may share a similar mechanism to the previous report [26].

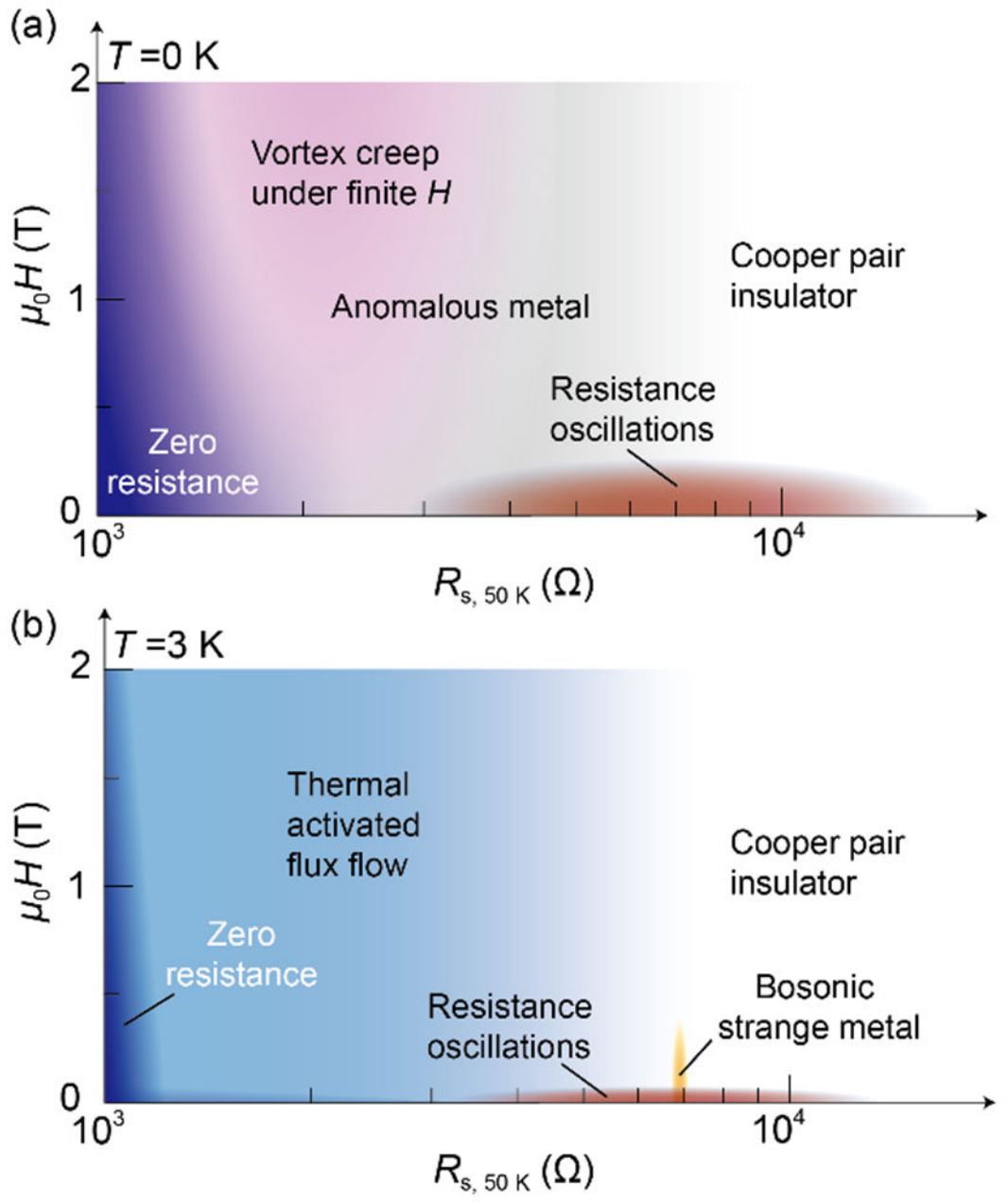

FIG. 5. Phase diagram of sample S1 at different temperatures. (a) Phase diagram at 0 K. The pristine sample exhibits zero resistance up to high magnetic fields (dark blue region), defined as the regime where the resistance is below the noise level. Upon etching, the resistance at high temperature ($R_{s,\ 50\ K}$) increases, and the dissipationless transport is suppressed under a small applied magnetic field. An AM state emerges at finite magnetic fields, which can be possibly attributed to quantum creep of vortices driven by the external magnetic field (pink region). As $R_{s,\ 50\ K}$ increases further, the AM state extends down to zero magnetic field, and resistance oscillations with a period of $h/2e$ appear (red region). Eventually, the system transitions into a Cooper pair insulator (white region). (b) Phase diagram at 3 K. At this elevated temperature, the system retains a similar initial state (zero resistance) and final state (Cooper pair insulator) as at 0 K. The intermediate states are affected by thermal fluctuations. They are replaced by thermal activated flux flow (light blue area) and bosonic strange metal state (orange area). Resistance oscillations persist in the red-shaded region. All phase boundaries in both panels are estimated based on the experimental data of sample S1 presented in Figs 1–4.

## III. SUMMARY

The phase diagram of sample S1 in Fig. 5 summarizes our results. We use $R_{s,\ 50\ K}$ as the $x$-axis, which represents the increasing disorder and superconducting fluctuation level upon etching. In the pristine sample, vortices remain pinned up to very high magnetic fields, resulting in zero resistance at both 0 K and 3 K (indicated in dark blue on the left side of Figs. 5(a) and (b)). Notably, in another sample with the same stoichiometry, we observed zero resistivity under a magnetic field of 35 T at 1.6 K. Patterning the film into a superconducting network enhances superconducting fluctuations and suppresses the superconductivity. We detect MR oscillations with a period of $h/2e$ in red regions of Figs. 5(a) and (b), which is clear evidence that Cooper pairs participate in the electrical transport although zero resistivity is not obtained. We observe multiple phases at different fluctuation strengths and temperatures. At zero temperature, the system exhibits AM states. The AM state in the pink region arises when the applied magnetic field drives the system out of the dissipationless regime, possibly associated with magnetic field induced quantum creep of vortices. The AM state in the gray region at a higher disorder strength persists down to zero magnetic field, which is possibly a result of Ohmic dissipation originating from the coupling of bosonic and fermionic modes. At 3 K (Fig. 5(b)), thermal activation becomes significant. The finite resistance in the light blue area is attributed to thermally activated flux flow. With further enhancement of disorder strength, a bosonic strange metal state emerges (orange region in Fig. 5(b)), characterized by linearly temperature-dependent resistance. Finally, under sufficiently strong disorder, the system is driven into a Cooper pair insulator. The sequence of phase transitions demonstrates rich and tunable states governed by phase coherence in nickelate superconductors.

Our work on nanopatterned samarium nickelate is significant not only for advancing the understanding of the bosonic metallic phase but also for deepening insight into the mechanisms of nickelate superconductors. The bosonic metallic states observed in nanopatterned nickelates exhibit several distinctions from previously studied superconducting networks [24-27]. Earlier works have only reported AM states persisting down to zero magnetic field in superconducting networks. However, depending on the etching time, we observe two types of AM states in our samples: one that emerges from a dissipationless state under an applied magnetic field, and another that persists down to zero field. The former behavior—possibly associated with quantum vortex creep—has not been reported in similar superconducting networks [24-27]. Moreover, both the bosonic strange metal and AM states that are present at zero magnetic field in nickelate samples seem to be more sensitive to the magnetic field. The linear $R_s(T)$ and the resistance plateau can be suppressed at a relatively small magnetic field [25, 26]. These distinctions suggest possible different vortex dynamics in nickelate superconductors, and a microscopic model that reflects the material-specific properties of nickelates is essential for understanding the bosonic metallic states in our samples. In terms of the superconducting mechanism of nickelates, our results provide direct evidence for $2e$ pairing, offering important constraints on theoretical models of nickelate superconductivity. Technically, this work represents the first study of nanofabricated superconducting devices based on nickelates, providing valuable insights for future investigations into key questions such as pairing symmetry.

## DATA AVAILABILITY

The raw data in the current study are available from the corresponding author upon reasonable request.

## APPENDIX: METHODS

### A. Growth and topotactic reduction of the $Sm_{0.95-x}Eu_xCa_{0.05}NiO_2$ films

10-nm-thick $Sm_{0.95-x}Eu_xCa_{0.05}NiO_3$ ($x$=0.2, 0.22, 0.24) films were epitaxially deposited on $(LaAlO_3)_{0.3}(Sr_2TaAlO_6)_{0.7}$ (001) substrates via pulsed laser deposition (KrF excimer laser, $\lambda$ = 248 nm) at 600 °C under 100 mTorr oxygen pressure with a fluence of 2.6 J/cm² and a repetition rate of 3 Hz. Subsequently, an epitaxial ~2-nm-thick $SrTiO_3$ capping layer was deposited using a lower fluence of 0.6 J/cm². Topotactic reduction was carried out in a vacuum chamber using $CaH_2$ powders. Reduction progress was monitored in real-time via two-probe resistance measurements, with the process terminated upon reaching the resistance minimum. Reduction temperatures (thermocouple-measured) ranged from 270 °C to 310 °C, with total annealing times of 0.5–1.5 hours.

### B. Fabrication of the periodic $Sm_{0.95-x}Eu_xCa_{0.05}NiO_2$ networks

To fabricate the periodic nanostructures, AAO masks were transferred onto $Sm_{0.95-x}Eu_xCa_{0.05}NiO_2$ films followed by RIE. The AAO masks coated with polymethyl methacrylate (PMMA) were purchased from TopMembranes Technology Co., Ltd. The 200 nm thick AAO masks have a triangular array of holes with 50 nm diameter and 100 nm center-to-center distance. To prepare the masks for transfer, they were cut to appropriate sizes and immersed in acetone to remove the PMMA layer. While suspended in the acetone, the masks were floated onto the surface of the $Sm_{0.95-x}Eu_xCa_{0.05}NiO_2$ samples. Upon drying in air, the masks adhered to the film surface. For the RIE process, different gas mixtures and powers were employed for different samples. Sample S1, S3, and S4 were etched using an Ar/$CF_4$ mixture (20 sccm/20 sccm) at 5 Pa and 20 W, while Sample S2 was etched using pure Ar (20 sccm) at 30 Pa and 300 W.

### C. Transport measurements

Electrical contacts were fabricated by cold-pressing indium cubes onto the films after selectively removing the AAO mask at the contact areas. The same contacts were used for transport measurements after each etching step. A standard four-probe configuration was employed for resistance measurements. Unless otherwise specified, lock-in techniques were used with an excitation current of 200 nA for S1#3-#5, S3, and 500 nA for the rest samples. Transport measurements were performed in two different cryostats: one equipped with a $^3He$ insert with a base temperature of 0.26 K (used for S1#2–#5), and another without the insert with a base temperature of 1.6 K (used for S1#0, #1, and S2-S4). Low-pass RC filters with a cutoff frequency of 10 kHz were installed on each wire.

### D. XRD measurements

XRD measurements were performed using an automated X-ray diffractometer (SmartLab, Rigaku Corporation). To minimize oxidation, the measurements of the network were carried out immediately after the sample was removed from the cryostat. Prior to characterization, the AAO mask was removed using Scotch tape.

## ACKNOWLEDGEMENTS

We thank Haiwen Liu for helpful discussions. This work is financially supported by National Natural Science Foundation of China (92565102 (M.L.), 92165104, 12074038 (K.C.), 12504161 (H.W.), 12504166 (Z.G), 92265106 (J.C.), 12304189 (P.D.)); National Key R&D Program of China (2024YFA1408101, 2022YFA1403101 (Z.C.), 2024YFA1409001 (J.C.)); Quantum Science and Technology-National Science and Technology Major Project (2023ZD0300500 (K.C.), 2021ZD0302403 (J.C.), 2023ZD0300502 (P.D.)); Beijing Municipal Science & Technology Commission (Z221100002722013 (K. C.)); Scientific Research Innovation Capability Support Project for Young Faculty (ZYGXQNJSKYCXNLZCXM-D5 (K.C.)); International Station of Quantum Materials (Z.C.); Beijing Natural Science Foundation (1232035 (P.D.)). The work performed at City University of Hong Kong was supported by the National Natural Science Foundation of China (12174325), a Guangdong Basic and Applied Basic Research Grant (2023A1515011352), and by research grants from the Research Grants Council (RGC) of the Hong Kong Special Administrative Region, China, under Early Career Scheme, General Research Fund and ANR-RGC Joint Research Scheme (CityU 21301221, CityU 11309622, CityU 11300923, A-CityU102/23 and CUHK 24306223). Part of the work utilised the equipment support through a Collaborative Research

Equipment Grant from RGC (C1018-22E).